\documentclass[aps,prl,groupedaddress,superscriptaddress,preprint,floatfix]{revtex4-1}

\usepackage{amsmath,amssymb}
\usepackage{booktabs}
\usepackage{graphicx}
\usepackage{xcolor}
\usepackage{bm}
\usepackage{mathtools}
\mathtoolsset{showonlyrefs=true}

\begin{document}

\title{Magnetic Monopole Supercurrent through a Quantum Spin Ice Tunnel Junction}

\author{Sho Nakosai}
\affiliation{Condensed Matter Theory Laboratory, RIKEN, Wako, Saitama 351--0198, Japan.}
\author{Shigeki Onoda}
\affiliation{Condensed Matter Theory Laboratory, RIKEN, Wako, Saitama 351--0198, Japan.}
\affiliation{Quantum Matter Theory Research Team, RIKEN Center for Emergent Matter Science (CEMS), Wako, Saitama 351--0198, Japan.}

\begin{abstract}
Magnetic monopoles are hypothetical particles that may exist as quantized sources and sinks of the magnetic field. In materials, they may appear in an emergent quantum electrodynamics described by a U(1) lattice gauge theory. Particularly, quantum spin ice hosts monopoles as bosonic spinons coupled to emergent gauge fields in a U(1) quantum spin liquid, namely, a deconfined Coulomb phase. When monopoles are condensed to form a long-range order, monopoles and gauge fields are screened and confined. Here we show, however, that monopole supercurrent flows across a junction of two ferromagnets that are weakly linked through and placed on top of the U(1) QSL, when a gauge-invariant phase difference of spinons across the junction is generated by quenching or an applied electric voltage parallel to the junction. This novel phenomenon paves the way to a new paradigm of spinonics for a dissipationless control of magnetism.
\end{abstract}

\maketitle

A gauge symmetry and its spontaneous symmetry breaking are key concepts in modern physics for unifying theories of physical phenomena that occur at different energy scales and in different phases. Electrical superconductivity~\cite{nambu:60,anderson:63}, a separation of weak and electromagnetic interactions~\cite{weinberg:QuantumTheoryOfFields}, and a possible color superconductivity~\cite{alford:08} are ascribed to spontaneously broken U(1), SU(2)$\times$U(1), and SU(3) gauge symmetries, respectively, in the standard model. In condensed matter, a gauge symmetry may spontaneously appear with a deconfinement of fractionalized quasiparticles coupled to emergent gauge fields, when an intrinsic topological order is realized by many-body effects at low temperatures~\cite{wen:QuantumFieldTheory}, as in fractional quantum Hall states~\cite{laughlin:83,moore:91,wen:QuantumFieldTheory}. Of our interest is yet another prototype found in quantum spin liquids (QSLs).

QSLs~\cite{wen:02,lee:08,balents:10} are long-range entangled, topological states of magnetic materials, where quantum fluctuations and a geometrical frustration of magnetic interactions prevent electronic spins from showing any spontaneous symmetry-breaking order at zero temperature. They host elementary excitations of nearly free spinons, namely, matters, carrying fractional spin quanta and of emergent (gapless) compact gauge fields (for continuous gauge groups)~\cite{fradkin:79}. Thus, they are described as deconfined phases of associated compact gauge theories.
On the other hand, conventional ordered phases with spontaneously broken symmetries are described as Higgs phases.
In this case, a Higgs field expressed by (multiple) spinons acquires a macroscopic phase coherence, although the macroscopic phase is not observable since it is not gauge-invariant.
In particular, when the Higgs field carries the fundamental gauge charge, gauge charges and fields are perfectly screened and thus confined~\cite{fradkin:79}. These charge-1 Higgs confined phases contain conventional spin waves as elementary excitations but not spinons.

Now let us consider the case where two Higgs confined phases are weakly linked through a deconfined Coulomb phase that can host deconfined gauge fields. Then, a finite gauge-invariant phase difference of spinons can be created through the link, and one may expect an analogous Josephson effect: tunneling supercurrent of the gauge charge carried by spinons can be generated across the junction. The goal of this paper is a theoretical demonstration of this phenomenon in candidate systems.

Remarkably, quantum spin ice (QSI) modeled for both magnetic rare-earth pyrochlores~\cite{molavian:07,onoda:10,ross:11,chang:11,kimura:13,takatsu:15} and $A$-site deintercalated spinel iridates Ir$_2$O$_4$~\cite{onoda:16} has recently been highlighted as a unique laboratory for hosting a bosonic U(1) QSL~\cite{hermele:04,motrunich:05} and charge-1 Higgs confined phases in a U(1) lattice gauge theory~\cite{hermele:04,savary:12,lee:12}.
It is described by the following Hamiltonian~\cite{onoda:10},
\begin{align}
	H=\frac{J}{2}\sum^{n.n.}_{\langle \bm{r},\bm{r}^\prime\rangle}&\Bigg[
		\left(S^z_{\bm{r}}S^z_{\bm{r}'}+\frac{1}{4}\right)
		+ \delta S^+_{\bm{r}}S^-_{\bm{r}^\prime}
		+ qe^{2i\phi_{\bm{r},\bm{r}'}} S^+_{\bm{r}}S^+_{\bm{r}^\prime}
		\nonumber\\[-5pt]
		&+ Ke^{i\phi_{\bm{r},\bm{r}'}} \left(S^{+}_{\bm{r}}S^{z}_{\bm{r}'}+S^{z}_{\bm{r}}S^{+}_{\bm{r}'}\right)
	\Bigg] + \mathrm{h.c.},
	\label{eq:H:spin}
\end{align}
with a spin-ice-rule coupling $J$ and three dimensionless nearest-neighbor exchange coupling constants $(\delta,q,K)$, where $\bm{S}_{\bm{r}_\mu}=\sum_{i=x,y,z}S^i_{\bm{r}_\mu}\bm{e}^i_\mu$ stands for a pseudospin-$1/2$ operator at a pyrochlore lattice site $\bm{r}_\mu$ of a sublattice index $\mu=0,\dots,3$. The $C_2$-invariant set of $\mu$-dependent local coordinates $(\bm{e}^x_\mu,\bm{e}^y_\mu,\bm{e}^z_\mu)$ (Fig.~1(a)) and the bond-dependent phase factor $\phi_{\bm{r},\bm{r}'}$ are given in Table~S1~\cite{supplement}.

\begin{figure}
  \centering
	\includegraphics[width=.45\hsize]{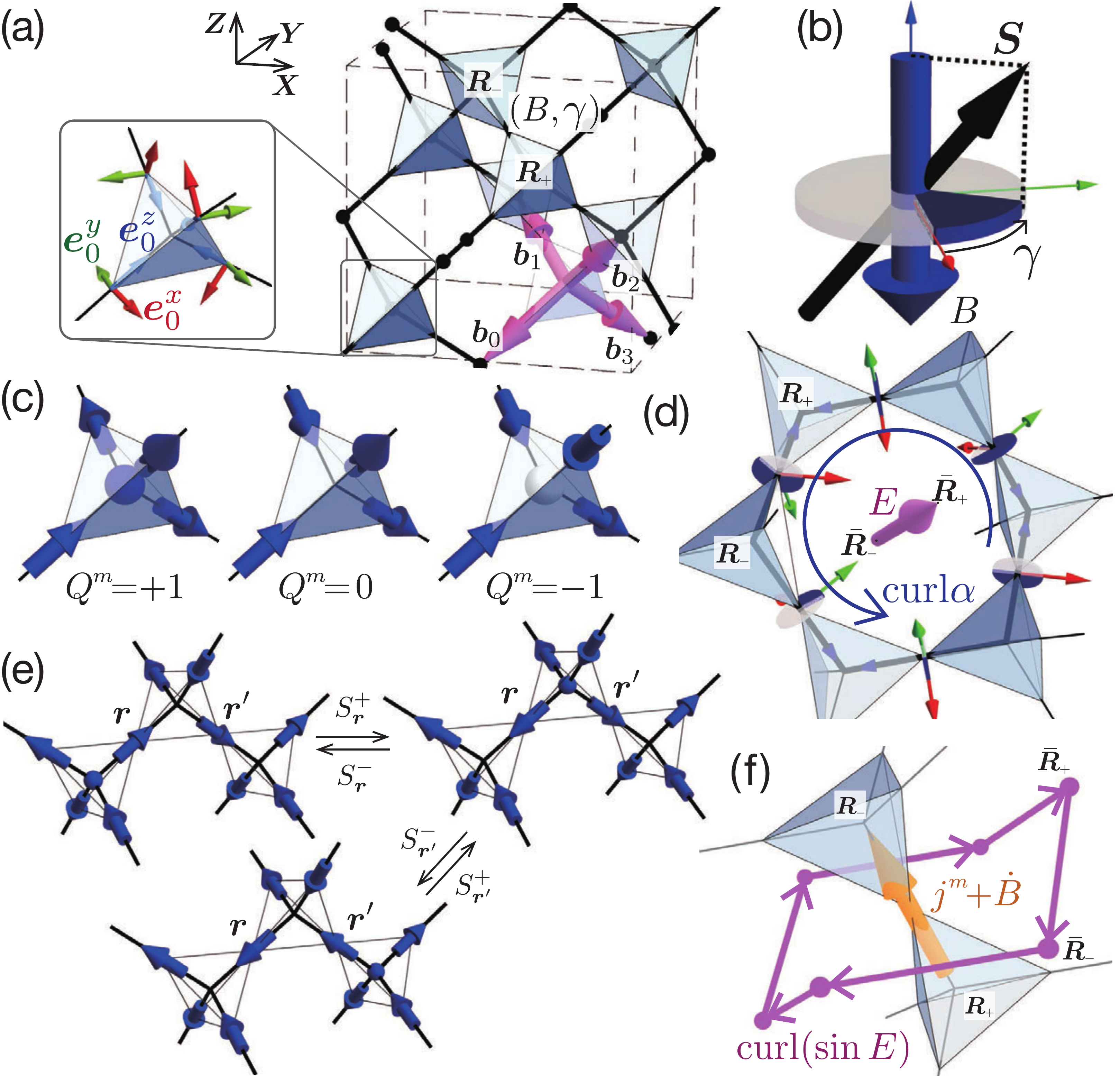}
	\caption{\label{fig:pyrochlore}
    (a) Pyrochlore lattice comprising a corner-sharing network of tetrahedra. Directed variables $(B,\gamma)$ are defined on the links (bold lines) of the diamond lattice sites that represent the centers of tetrahedra.
    The inset shows the local coordinates $(\bm{e}^x_\mu,\bm{e}^y_\mu,\bm{e}^z_\mu)$.
    (b) Representation of $\bm{S}$ by $B$ (blue arrow) and $\gamma$.
    (c) Configuration of magnetic monopole charge $Q^{\mathrm{m}}$ constructed from Eq.~\eqref{eq:Q}. $Q^{\mathrm{m}}=+1/-1$ is presented by the blue/white ball.
    (d) Lattice curl of $\alpha$ and the associated $E$.
    (e) Hopping processes of a magnetic monopole.
  (f) The modified Faraday law, relating $j^{\mathrm{m}}$ and the time derivative $\dot{B}$ on a diamond lattice link to a lattice curl of $\sin E$ along the dual diamond lattice links.}
\end{figure}

Equation~\eqref{eq:H:spin} has been mapped onto a compact U(1) lattice gauge theory~\cite{hermele:04,moessner:06,savary:12,lee:12} as follows.
A canonical conjugate pair of the analogous magnetic field $B$ and the azimuthal angle $\gamma$ is introduced through $S^z_{\bm{r}_\mu}=\sigma B_{\ell_{\bm{r}_\mu,\sigma}}$ and $S^\sigma_{\bm{r}_\mu}=\exp[i\gamma_{\ell_{\bm{r}_\mu,\sigma}}]$ (Fig.~1(b)). These fields are directed variables defined at the link $\ell_{\bm{r}_\mu,\sigma}: \bm{R}_\sigma\to\bm{R}_\sigma+\sigma\bm{b}_\mu$ from the diamond lattice site $\bm{R}_\sigma$ located at the center of an upward/downward-oriented ($\sigma=+/-$) tetrahedron to the nearest-neighbor diamond lattice site $\bm{R}_\sigma+\sigma\bm{b}_\mu$ through the pyrochlore site $\bm{r}_{\mu}=\bm{R}_{\sigma}+\sigma\bm{b}_{\mu}/2$ (Fig.~1(a)). We can write $\gamma_{\ell_{\bm{r}_\mu,\sigma}} \allowbreak =\varphi_{\bm{R}_\sigma+\sigma\bm{b}_\mu} \allowbreak -\varphi_{\bm{R}_\sigma} \allowbreak + \alpha_{\ell_{\bm{r}_\mu,\sigma}}$.
Here, $\varphi$ stands for a phase of a spinon (magnetic monopole) field $\Phi_{\bm{R}_\sigma}=\exp[i\varphi_{\bm{R}_\sigma}]$.
It satisfies the canonical conjugate relation $\left[Q^{\mathrm{m}}_{\bm{R}},\varphi_{\bm{R}'}\right]=i\delta_{\bm{R},\bm{R}'}$ with the quantized analogous magnetic monopole charge (Fig.~1(c)),
\begin{align}
  Q^{\mathrm{m}}_{\bm{R}_\sigma}&=(\mathrm{div}B)_{\bm{R}_\sigma}
	\equiv\sum_{\mu=0}^3B_{\ell_{\bm{r}_\mu,\sigma}} \,\, \mbox{(magnetic Gauss law)}.
	\label{eq:Q}
\end{align}
We have also introduced the dual vector potential $\alpha$ that is canonical conjugate to $B$;
$\left[\alpha_{\ell_{\bm{r}_\mu,\sigma}}, B_{\ell_{\bm{r}'_\nu,\sigma'}}\right] = i\sigma\sigma'\delta_{\bm{r}_\mu,\bm{r}'_\nu}$.
An analogous electric field is defined as $E_{\bar{\ell}_{\bar{\bm{r}}_\mu,\sigma}} \allowbreak={(\mathrm{curl}\alpha)}_{\bar{\ell}_{\bar{\bm{r}}_\mu,\sigma}} \allowbreak -E^0$ $\bmod$ $2\pi$,
at each link $\bar{\ell}_{\bar{\bm{r}}_\mu,\sigma}$ from the dual diamond lattice site $\bar{\bm{R}}_\sigma$ to its nearest-neighbor $\bar{\bm{R}}_\sigma+\sigma\bm{b}_\mu$ through the dual pyrochlore lattice site, $\bar{\bm{r}}_{\mu}=\bar{\bm{R}}_\sigma+\sigma\bm{b}_\mu/2$, located at the center of the hexagonal plaquette. Henceforth, $\mathrm{curl}\alpha$ stands for the directed sum of $\alpha$ along the hexagonal plaquette around the link $\bar{\ell}_{\bar{\bm{r}}_{\mu},\sigma}$ (Fig.~1(d)). We have also introduced a uniform shift $E^0=0$ for $\delta<0$ and $\pi$ for $\delta>0$.
An analogous electric monopole charge is given by
\begin{align}
	Q^{\mathrm{e}}_{\bm{R}_\sigma}=\frac{1}{2\pi}(\mathrm{div}E)_{\bm{R}_\sigma} \,\, \mbox{(electric Gauss law)}
	\label{eq:Qe}
\end{align}
and is quantized to integer values. Variables introduced in this paper and their actions under symmetry operations are summarized in Table~S2~\cite{supplement}.
In particular, $\sin E$ linearly couples to the real electric field and polarization by symmetry as is obtained from  a gauge-invariant generalization of Ref.~\onlinecite{PhysRevB.96.125145}.

The spin-ice-rule interaction bears a finite excitation energy for creating a pair of oppositely charged magnetic monopoles from a spin-ice vacuum satisfying $Q^{\mathrm{m}}_{\bm{R}}\!=\!0$, as in the classical case ($\delta,q,K=0$)~\cite{castelnovo:08}. Then, an analogous quantum electrodynamics (QED) without matter fields can be formulated by integrating out spinons and then performing a duality mapping and a noncompactification~\cite{hermele:04}. This formalism is, however, inconvenient for considering effects of a Bose-Einstein condensation of spinons explicitly. For our purpose of discussing interfaces between the U(1) QSL and charge-1 Higgs confined phases, an analogous compact QED will be reformulated below by keeping all the degrees of freedom of magnetic and electric monopoles and gauge fields.

As long as magnetic monopole excitations are gapped, as is the case for small $|\delta|$, $|q|$ and $|K|$, the offdiagonal terms with respect to the number $N^{\mathrm{m}}=(1/2)\sum_{\bm{R}}|Q^{\mathrm{m}}_{\bm{R}}|$ of magnetic monopole pairs can be treated as perturbations. This leads to the frustrated compact U(1) lattice gauge theory described by the Hamiltonian $H_{\mathrm{LGT}}=H_{\mathrm{m}}[Q^{\mathrm{m}}]+H_{\mathrm{PLGT}}[E,B]+H_{\mathrm{matter}}[\Phi,\Phi^\dagger,e^{i\alpha},B]$ in the restricted subspace that is diagonal with respect to $N^{\mathrm{m}}$.
Here, $H_{\mathrm{m}}[Q^{\mathrm{m}}]=\allowbreak (J/2) \allowbreak \sum_{\bm{R}}{Q^{\mathrm{m}}_{\bm{R}}}^2 \allowbreak -(J_B/3)\sum_{\langle\bm{R},\bm{R}'\rangle}^{n.n.}Q^{\mathrm{m}}_{\bm{R}}Q^{\mathrm{m}}_{\bm{R}'}$ gives a magnetic monopole charge potential, and
\begin{align}
	H_{\mathrm{PLGT}} =
  -J_{E}\!\!\sum_{\bar{\bm{r}}_\mu} \cos E_{\bar{\ell}_{\bar{\bm{r}}_{\!\mu},\!+}}
  -J_{B}\sum_{\bm{r}_{\mu},\!\nu\ne\mu} B_{\ell_{\bm{r}_{\!\mu},\!+}} B_{\ell_{\bm{r}_{\mu}+\bm{b}_{\nu}-\bm{b}_{\mu},\!+}},
  \label{eq:H_PLGT}
\end{align}
is a pure lattice gauge theory Hamiltonian with the coupling constants for the hexagonal ring exchange and second-neighbor magnetic interactions, $J_E\sim (3/2)J|\delta|^3$ and $J_B\sim (3/2)JK^2$, respectively.
Lastly, $H_{\mathrm{matter}}=\sum_{\bm{R},\bm{R}'}\Phi^\dagger_{\bm{R}}h^{\mathrm{matter}}_{\bm{R},\bm{R}'}[e^{i\alpha},B]\Phi_{\bm{R}'}$ is the matter part treated in Refs.~\onlinecite{savary:12,lee:12}. It is given by
\begin{align}
	&\frac{J}{2}\sum_{\sigma,\bm{R}_\sigma,\mu\ne\nu} \left[
		\delta\Phi^\dagger_{\bm{R}_\sigma+\sigma\bm{b}_\mu}
		e^{-i(\alpha_{\ell_{\bm{r}_\mu,\sigma}}-\alpha_{\ell_{\bm{r}_\nu,\sigma}})}
		\Phi_{\bm{R}_\sigma+\sigma\bm{b}_\nu}
		\right.
		\nonumber \\[-5pt]
		&+\left.
		K\sigma B_{\ell_{\bm{r}_\mu,\sigma}}
		\left(\Phi_{\bm{R}_\sigma}^\dagger
		e^{i(\alpha_{\ell_{\bm{r}_\nu,\sigma}}+\sigma\phi_{\mu\nu})}
		\Phi_{\bm{R}_\sigma+\sigma\bm{b}_\nu} + \mathrm{h.c.}\right)
	\right],
	\label{eq:hopping}
\end{align}
which describes gauge-invariant first- and second-neighbor hopping terms of magnetic monopole, as depicted in Figs.~1(e). (Correlated hopping terms proportional to $q$ have been left out since they give irrelevant perturbations as far as $|q|$ is small~\cite{lee:12}.)
Now monopole current can be defined as
$j^{\mathrm{m}}_{\ell}=\partial H_{\mathrm{matter}}/\partial \alpha_{\ell}$. It exactly satisfies the conservation law of the monopole charge,
\begin{align}
	\dot{Q}^{\mathrm{m}}_{\bm{R}_\sigma}+(\mathrm{div}j^{\mathrm{m}})_{\bm{R}_\sigma}=0.
	\label{eq:conservation}
\end{align}
This $j^{\mathrm{m}}$ generates $E$ through
\begin{align}
	-J_E(\mathrm{curl}(\sin E))_{\ell}
	=\dot{B}_{\ell}+j^{\mathrm{m}}_{\ell} \,\, \mbox{(modified Faraday law)}.
	\label{eq:Faraday}
\end{align}
(See Fig.~1(f).) Similarly, it is ready to derive
\begin{align}
	U_{\mathrm{B}}(\mathrm{curl}B)_{\bar{\ell}}=\dot{E}_{\bar{\ell}}+j^{\mathrm{e}}_{\bar{\ell}} \,\, \mbox{(Maxwell-Amp\'ere law)},
	\label{eq:Ampere}
\end{align}
with a Lagrange multiplier $U_B$ assuring the constraint $B^2=1/4$ and the analogous conserved electric current $j^{\mathrm{e}}_{\bar{\ell}}=-(\mathrm{curl}(\partial H_{\mathrm{LGT}}/\partial B))_{\bar{\ell}}$ that linearly contributes to a real magnetic toroidal moment.

The high-temperature symmetry-unbroken phase of this theory is given by a thermally confined phase where the classical spin ice physics~\cite{castelnovo:08} is effective~\cite{supplement}. In this case, thermal fluctuations of $E$ wash out sine terms in Eq.~(\ref{eq:Faraday}).
On cooling below $J_E$, fluctuations of $E$ and $Q^{\mathrm{e}}$ become suppressed. Then, Eq.~(\ref{eq:Faraday}) can be linearized in $E$, forming the set of analogous Maxwell equations. This leads to deconfined Coulomb phases at zero temperature. There exist two distinct Coulomb phases, namely, a bosonic U(1) QSL~\cite{hermele:04,savary:12,lee:12} and a Coulomb ferromagnet~\cite{savary:12}, where $B$ averages to zero and a finite value, as expected for $J_B\ll J_E$ and $J_B\gg J_E$, respectively. These Coulomb phases can host $E$ fields and analogous gapless photons as well as gapped magnetic and electric monopole excitations~\cite{hermele:04}, behaving as monopole insulators.

With increasing $|\delta|$, $|q|$ or $|K|$, magnetic monopole excitations become softened and eventually Bose-Einstein condensed. This spontaneously breaks the global U(1) gauge symmetry with $\gamma$ being fixed, leading to charge-1 Higgs confined phases with symmetry-breaking long-range orders of $(S^x_{\bm{r}},S^y_{\bm{r}})$~\cite{savary:12,lee:12}.
Then, spinon fields in $H_{\mathrm{matter}}$ are decomposed as $\Phi_{\bm{R}} \allowbreak =\left( \sqrt{n_{\bm{R}}} + \Delta\Phi_{\bm{R}} \right) e^{i\varphi_{\bm{R}}}$,
where $\langle\Phi\rangle=\sqrt{n}e^{i\varphi}$ gives the condensed part with the amplitude $\sqrt{n}$ and the phase $\varphi$ and $\Delta\Phi$ is a real scalar Higgs field.
Magnetic and electric monopole excitations are eliminated from the spectrum since their phases are absorbed into gauge-invariant quantities. Thus, the theory can be expanded in the fluctuating part of $\gamma$. Accordingly, the gauge photon modes turn into (pseudo-)Nambu-Goldstone modes that remain gapless for $K=q=0$ and are gapped otherwise by the Higgs mechanism~\cite{nambu:60,anderson:63b,higgs:64}. In the gapped case, there appears an analogous dual Meissner effect: $E$ is shielded by monopole supercurrent and exponentially decays from a surface into the bulk. Such ferromagnet, henceforth dubbed a Higgs ferromagnet (HFM), has been reported in Yb$_2$Ti$_2$O$_7$~\cite{chang:11}, which shows a nearly collinear ferromagnetic long-range order with a [001] uniform magnetization.
See Table~S2~\cite{supplement} for a summary of properties of the possible phases.

Now we consider a junction device of QSI systems.
Two HFMs are weakly linked through the U(1) QSL, as shown in Fig.~2(a) where the blue and grey shaded regions correspond to HFMs and a U(1) QSL, respectively.
The two interfaces of the U(1) QSL with the left and right HFMs are taken to be normal to the crystallographic $\bm{Z}=[001]$ direction.
For simplicity, we assume that effects of $\sin E$ on $j^{\mathrm{m}}$ are negligibly small, as in dual type-II superconductors. Then, the gauge fields $(e^{i\alpha_{\ell_{\bm{r}_\mu,\sigma}}}, B_{\ell_{\bm{r}_\mu,\sigma}})$ can be treated as real variational parameters $(\chi_{\ell_{\bm{r}_\mu,\sigma}}, \langle B_{\ell_{\bm{r}_\mu,\sigma}}\rangle)$. In particular, to minimize $H_{\mathrm{PLGT}}$, we take $E\!=\!0$ as in the gauge mean-field theory~\cite{savary:12,lee:12}, and adopt a gauge choice where $e^{i\alpha}$ is real and positive everywhere.
In this frozen gauge-field approximation, the Lagrangian is expressed as
\begin{align}
  L&{}= \sum_{\sigma,\,\bm{R}_{\sigma}} \frac{1}{2J}|\Delta\dot{\Phi}_{\bm{R}_{\sigma}}|^2
	- \sum_{\bm{R},\bm{R}^{\prime}} \Phi_{\bm{R}}^{\ast} h^{\mathrm{matter}}_{\bm{R}\bm{R}^{\prime}}[\chi,\langle B\rangle] \Phi_{\bm{R}^{\prime}}
	\nonumber \\[-5pt]
	-&{} \sum_{\sigma, \bm{R}_\sigma}\frac{u_{\bm{R}_{\sigma}}^{\Phi}}{2} {\left( |\Phi_{\bm{R}_{\sigma}}|^{2}-1 \right)}^{2}
	-\frac{J}{2}\sum_{\sigma,\, \bm{R}_\sigma}{\left[\sum_\mu\langle B_{\ell_{\bm{r}_\mu,\sigma}}\rangle\right]}^2
	\nonumber \\[-5pt]
	-&{} \sum_{\sigma,\, \mu,\, \ell_{\bm{r}_\mu,\sigma}} \frac{\lambda_{\ell_{\bm{r}_\mu,\sigma}}^{g}}{2}
	\left(
	\langle B_{\ell_{\bm{r}_\mu,\sigma}}^{2} \rangle
	+ \chi_{\ell_{\bm{r}_\mu,\sigma}}^{2}- \frac{1}{4}
	\right),
  \label{eq:l}
\end{align}
with the Lagrange multipliers $\lambda_{\ell_{\bm{r}_\mu,\sigma}}^{g}$ and $u_{\bm{R}}^{\Phi}$ for treating the local constraints.
Note that the translational invariance along the $\bm{Z}$ direction is broken, in contrast to the bulk cases~\cite{savary:12,lee:12}.
The spatially dependent real variational parameters $(\chi, \langle B\rangle, \lambda^g, u^\Phi)$ and complex parameters $\langle\Phi\rangle$ are determined so that they satisfy the saddle point conditions with respect to themselves, except that the condition with respect to $u^\Phi$ is modified by Gaussian fluctuations of the Higgs field as
$\left\langle \middle| \Delta\Phi_{\bm{R}}\middle|^{2} \right\rangle = \int d\omega\,G_{\bm{R},\bm{R}}(\omega)f(\omega) = 1-n_{\bm{R}}$,
with the Bose-Einstein distribution function $f(\omega)$,
and the Green function $G(\omega)$ of the real scalar Higgs field where
$G^{-1}_{\bm{R}\bm{R}^{\prime}}(\omega) = 2\left( \frac{\omega^{2}}{2J} - u^{\Phi}_{\bm{R}} \right)\delta_{\bm{R}\bm{R}^{\prime}} - 2\mathrm{Re}(h^{\mathrm{matter}}_{\bm{R}\bm{R}^{\prime}}e^{-i(\varphi_{\bm{R}}-\varphi_{\bm{R}'})})$.
To be explicit, we adopt $\delta=\delta_{\mathrm{HFM}}=-0.3$, $K=K_{\mathrm{HFM}}=1.0$, and $q=0$ in the HFMs and $\delta=0.3\delta_{\mathrm{HFM}}$, $K=0.3K_{\mathrm{HFM}}$, and $q=0$ in the U(1) QSL\@. The boundary condition is chosen so that $n_{\bm{R}} = 1$ beyond the left and right boundaries of the system and the spatial profile of the phase $\varphi_{\bm{R}}$ there is consistent with the uniform collinear $[001]$ or $[00\bar{1}]$ magnetic moment distribution $\bm{S}_{\bm{r}}=\bm{S}^{\mathrm{L/R}}$. In particular, we consider the head-to-tail (Fig.~2(a)) and head-to-head (Fig.~2(b)) configurations, which are dubbed the $Z$I$Z$ and $Z$I$\bar{Z}$ junctions, respectively, of $\bm{S}^{\mathrm{L}}$ and $\bm{S}^{\mathrm{R}}$.
Then, we vary the gauge-invariant phase difference $\Delta\varphi=\varphi^{\mathrm{R}}-\varphi^{\mathrm{L}}+\int_{\mathrm{L}\to\mathrm{R}}\alpha_{\bm{s}}\cdot d\bm{s}$, where $\varphi^{\mathrm{L}/\mathrm{R}}$ stands for a representative spinon phase in the left/right HFM. (For the details of $\varphi^{\mathrm{L}/\mathrm{R}}$, see Supplemental Information~\cite{supplement}.)
We numerically compute the saddle-point solutions for several choices of $\Delta\varphi$, and then the expectation values of spins, $(\langle S^{\pm}_{\bm{r}_{\mu}}\rangle, \langle S^{z}_{\bm{r}_{\mu}} \rangle) = \left(\langle\Phi^{\dagger}_{\bm{r}_{\mu}\mp\bm{b}_{\mu}/2} \Phi_{\bm{r}_{\mu}\pm\bm{b}_{\mu}/2}\rangle \chi_{\ell_{\bm{r}_{\mu},\pm}}, \langle B_{\ell_{\bm{r}_{\mu},+}}\rangle\right)$, and tunneling monopole supercurrent per a cubic unit cell, $j^{\mathrm{m}}\allowbreak=\frac{1}{N_{\mathrm{QSL}}}\allowbreak\sum_\mu \allowbreak\sum_{\bm{r}_\mu\in\mathrm{QSL}}\allowbreak\frac{b_\mu^Z}{|\bm{b}_\mu|}\allowbreak\frac{\partial H_{\mathrm{matter}}}{\partial \alpha_{\ell_{\bm{r}_\mu,+}}}[\sqrt{n}e^{i\varphi},\sqrt{n}e^{-i\varphi},\chi,\langle B\rangle]$, with the number $N_{\mathrm{QSL}}$ of cubic unit cells in the U(1) QSL region.

\begin{figure*}
	\centering
  \includegraphics[width=.95\hsize]{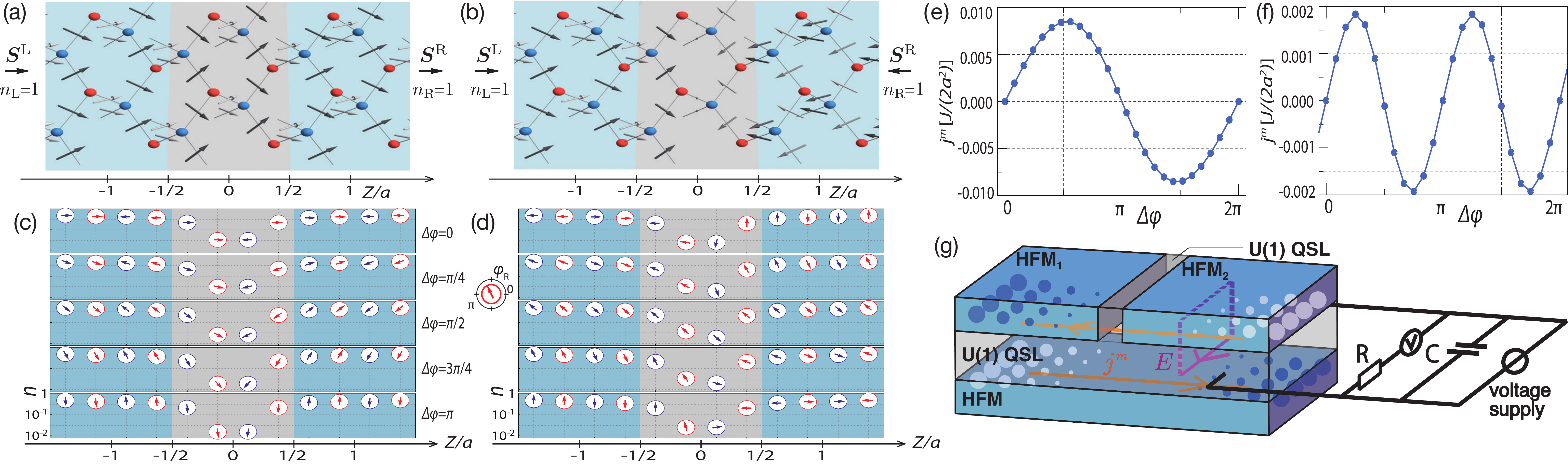}
	\caption{\label{fig:result}
	(a, b) Spin structures calculated for the $Z\mathrm{I}Z$ (a) and $Z\mathrm{I}\bar{Z}$ (b) junctions in the case of $\Delta\varphi=0$. Diamond lattice sites denoting centers of upward-oriented and downward-oriented tetrahedra are marked by balls colored in blue and red, respectively. Changes of the spin configurations during the cycle of $\Delta\varphi=0\to2\pi$ are indiscernible in the plot.
	(c, d) Spatial profiles of magnetic monopole condensates in the $Z\mathrm{I}Z$ (c) and $Z\mathrm{I}\bar{Z}$ (d) junctions for several choices of $\Delta\varphi$. The centers and directions of the compasses represent $n_{\bm{R}}$ and $\varphi_{\bm{R}}$, respectively.
  (e, f) Magnetic monopole supercurrent $j^{\mathrm{m}}$ along the $Z$ axis versus $\Delta\varphi$ for the $Z\mathrm{I}Z$ (e) and $Z\mathrm{I}\bar{Z}$ (f) junctions. The points represent the numerical solutions. The lines are guides to the eyes.
	(g) A QSI tunnel junction. Counter-propagating magnetic monopole supercurrents at the interfaces of the U(1) QSL buffer layer with the HFMs change the magnetic monopole charges (blue and white balls) at the edges of the interfaces.
}
\end{figure*}

Figures~2(a) and 2(b) presents the stable spin structures in the case of $\Delta\varphi=0$ for the $Z$I$Z$ and $Z$I$\bar{Z}$ junctions, respectively. $S^z$ and thus $B$ remain finite at the center $Z=0$ of the $Z$I$Z$ junction, whereas they average to zero in the $Z$I$\bar{Z}$ junction because of the incapability of satisfying the spin ice rule.
Results on magnetic monopole condensates in the $Z$I$Z$ and $Z$I$\bar{Z}$ junctions are shown in Figs.~2(c) and (d), respectively, for several choices of the gauge-invariant phase difference $\Delta\varphi$ from $0$ to $\pi$.
In the QSL region, the condensate fraction $n_{\bm{R}}$ as well as the amplitude of $\left( \langle S^{x}_{\bm{r}}\rangle, \langle S^{y}_{\bm{r}}\rangle \right)$ decays exponentially, and the condensate phase $\varphi_{\bm{R}}$ as well as the angle $\gamma_{\bm{r}}$ of $\left( \langle S^x_{\bm{r}}\rangle, \langle S^y_{\bm{r}}\rangle \right)$ deviates from the pattern of the bulk HFM and wind.
Accordingly, monopole tunneling current $j^{\mathrm{m}}$ appears as an oscillating function of $\Delta\varphi$ between the two HFMs, as shown in Figs.~2(e) and 2(f) in the case of $Z$I$Z$ and $Z$I$\bar{Z}$ junctions, respectively. While the periodicity of $j^{\mathrm{m}}$ in $\Delta\varphi$ is $2\pi$ in the $Z$I$Z$ junction, it is $\pi$ in the $Z$I$\bar{Z}$ junction.
This difference can be understood from  the Ginzburg-Landau analysis~\cite{supplement}, where the spatial profile of the phase of spinon condensates for the $Z$I$\bar{Z}$ junction does not allow a bilinear coupling of spinon fields between the left and right HFMs but quartic couplings.
We stress that monopole tunneling supercurrent can flow in the head-to-tail and head-to-head collinear configurations of magnetizations, where transverse spin current trivially vanishes.

In fact, monopole current must be accompanied by $\mathrm{curl}\sin E$ while $E$ cannot penetrate deeply into the HFMs. Besides, the $E$ field is not defined in the real vacuum, namely, it vanishes outside the device.
Therefore, the junction layer must be connected to a U(1) QSL buffer layer through a well controlled interface where the pyrochlore lattice structure of the junction layer matches that of the U(1) QSL buffer layer, so that the U(1) QSL buffer layer provides the analogous vacuum hosting the emergent gauge fields, namely, finite $\sin E$. Then, monopole current is allowed to flow at the interface between the junction and buffer layers.
If we assume $J\sim0.17$~meV~\cite{ross:11} in Yb$_2$Ti$_2$O$_7$, the magnitude of monopole current can be of the order of 1/ns/nm$^2$ at this interface, causing a magnetization reversal at the interface within $\sim$1~ns.
A setup for generating/detecting monopole current is depicted in Fig.~2(g).
The gauge-invariant phase difference ${\mit\Delta}\varphi$ can be generated and controlled by a real applied electric voltage between the two sides of the junction, since by symmetry, this induces $E$ and thus $\alpha$ that are screened in the HFM but not in the U(1) QSL\@. AC monopole tunneling supercurrent can also be generated by using an AC electric voltage.
When the bottom HFM layer is attached to the U(1) QSL buffer layer as shown in Fig.~2(g), a counter-propagating monopole supercurrent flows at the bottom interface, according to the analogous modified Faraday law (Eq.~\eqref{eq:Faraday}). Otherwise, i.e., when the QSL layer is exposed to the real vacuum, a dissipative monopole current may flow at the bottom if the $E$ field is interfered near the bottom surface and the surface hosts softened monopole excitations.
The counter-propagating monopole supercurrent at the two interfaces changes the monopole charge disproportionation at the ends of the interfaces. This change can be detected through a change in the real magnetic field outside the device. According to Ref.~\onlinecite{RevSciInstrum.87.093702}, their SQUID sensor provides the spatial resolution of a sub-micrometer. We can roughly estimate the change of the magnetic field to be of the order of $10^{-3}\Phi_0$ and thus is in principle measurable, where $\Phi_0$ is the elementary magnetic flux quantum, if we employ the area 10$^4$nm$^2$ of the side surface of the junction layer, the radius 100nm of SQUID, and the spacing 100nm between the sample and the detector.
Even without the applied electric voltage, quenching the two ferromagnets will pin the monopole phase in each ferromagnet randomly, and thus will induce monopole tunneling supercurrent. According to Eq.~\eqref{eq:Faraday}, this generates $E$ in the buffer layer, which can be probed as a real electric polarization or current.

An observation of magnetic monopole current provides compelling evidence of the emergent U(1) gauge fields and spinons, i.e., magnetic monopoles, in the QSI junction device. It can be used to examine whether the buffer layer hosts a U(1) QSL and whether the two weakly linked QSI materials host monopole condensates. Candidate materials include Pr$_2$Zr$_2$O$_7$ and Pr$_2$Ir$_2$O$_7$ for an electrically insulating and conducting U(1) QSL buffer layer, respectively, and Yb$_2$Ti$_2$O$_7$ for a QSI HFM\@.
Such device structure will be readily prepared layer by layer by means of the pulsed laser deposition method, which may validate the current minimal modeling of the junction. For example, an epitaxial growth of thin films of Dy$_2$Ti$_2$O$_7$ on Y$_2$Ti$_2$O$_7$ has been reported in Ref.~\onlinecite{NatCommun.5.3439}.
The monopole current may also appear between long-range ordered domains across short-range ordered domains in Yb$_2$Ti$_2$O$_7$ around the first-order ferromagnetic transition~\cite{chang:11}, leading to an enhancement of the dielectric constant.  Note that even when we replace HFMs with $Z_n$ QSLs, these interference phenomena occur but with a reduced periodicity in ${\mit\Delta}\varphi$ by $n$. The current theory can also be generalized to a junction of $Z_2$ QSL phases weakly linked through a fermionic U(1) QSL\@.

Spintronics for electric/magnetic controls of magnetic/electric degrees of freedom has attracted great attention for potential applications since the discovery of giant~\cite{fert:68,grunberg:86} and tunneling~\cite{yuasa:04,parkin:04} magnetoresistance phenomena. However, the fundamental quantity of transverse spin current $\bm{j}^{S^z}$ is not a conserved current but obeys
$\dot{S}^z=-\mathrm{div}\bm{j}^{S^z}+\bm{T}^z$
with a nonzero spin torque $\bm{T}^z$. This equation is understood in the analogous QED subject to Eq.~(\ref{eq:Faraday}) as we can assign $(\mathrm{div}\bm{j}^{S^z})_{\bm{r}_\mu}\propto j^{\mathrm{m}}_{\ell_{\bm{r}_\mu,+}}$ and $T^z_{\bm{r}_\mu}\propto(\mathrm{curl}(\sin E))_{\ell_{\bm{r}_\mu,+}}$.
An applied electric field can induce transverse spin current in the direction normal to the field. In this case, the induced magnetization change is normal to both the applied field and the direction of the current. Therefore, this effect can be easily distinguished from the phenomenon proposed in this paper.
Our theoretical proposal of tunneling supercurrent of magnetic monopoles and its coupling to a transverse electric polarization or current may open a novel paradigm of ``spinonics'', namely, using spinons for controlling magnetism by electric means for potential applications.
The dissipationless nature of the monopole supercurrent may solve drawbacks of conventional spintronics approaches in which (nearly) gapless excitations are manipulated.

\begin{acknowledgments}
  SO is grateful to M. Sigrist for enlightening discussions at an early stage of this work and to Y. Tokura, M. Kawasaki, N. Nagaosa, L. Balents, and E. Fradkin for useful conversations. The work was partially supported by Grants-in-Aid for Scientific Research under Grant No. 15H03692 from JSPS and under Grant No. 15H01025 from the MEXT of Japan  and by the RIKEN iTHES project. SN would like to acknowledge the support from the Motizuki Fund of Yukawa Memorial Foundation. Calculations were partly performed by using HOKUSAI supercomputers at RIKEN.
\end{acknowledgments}


\end{document}